# Spatio-temporal modelling of forest monitoring data: Modelling German tree defoliation data collected between 1989 and 2015 for trend estimation and survey grid examination using GAMMs

Nadine Eickenscheidt[1*], Nicole H. Augustin[2] and Nicole Wellbrock[1]

Instituts:

[1]*Thünen Institute of Forest Ecosystems, Alfred-Möller-Strasse 1, 16225 Eberswalde, Germany*

[2]*University of Bath, Claverton Down, Bath BA2 7AY, United Kingdom*

Email addresses:

Nadine.Eickenscheidt@lanuv.nrw.de
N.H.Augustin@bath.ac.uk
nicole.wellbrock@thuenen.de

Corresponding author:

*now at: State Agency for Nature, Environment and Consumer Protection of North Rhine-Westphalia, Leibnizstrasse 10, 45659 Recklinghausen, Germany

## Running title

Spatio-temporal modelling of German tree defoliation data

# Abstract

Spatio-temporal modelling of tree defoliation data of the German forest condition survey is statistically challenging, particularly due to irregular grids. In the present study generalized additive mixed models were used to estimate the spatio-temporal trends of defoliation of the main tree species from 1989 to 2015 and to examine the suitability of different monitoring grid resolutions. Although data has been collected since 1989, this is the first time the spatio-temporal modelling for entire Germany has been carried out. Besides the space-time component, stand age showed a significant effect on defoliation. The mean age and the species-specific relation between defoliation and age determined the general level of defoliation whereas fluctuations of defoliation were primarily related to weather conditions. The study indicates a strong association between drought stress and defoliation of all four main tree species. Intensity and duration of increased defoliation following drought stress was tree species-specific. Besides direct effects of weather conditions, indirect effects seem to play a further role. Defoliation of the comparably drought-tolerant species pine and oak was primarily affected by insect infestations following drought whereas considerable time for regeneration was required by beech following drought stress and recurring substantial fructification. South-eastern Germany has emerged as the region with the highest defoliation since the drought year 2003. This region was characterized by the strongest water deficits in 2003 compared to the long-term reference period. The present study gives evidence that the focus has moved from air pollution to climate change. Furthermore, the spatio-temporal model was used to carry out a simulation study to compare different survey grid resolutions. This grid examination indicated that an 8 x 8 km grid instead of the standard 16 x 16 km grid is necessary for spatio-temporal trend estimation and for detecting hot-spots in defoliation in space and time, especially regarding oak.

# 1 Introduction

The forest condition survey represents a fundamental part of the Europe-wide forest monitoring, which was initiated as consequence of the discussion on forest dieback in the 1980ies. Anthropogenic air pollution, in particular of sulphur (S) and nitrogen (N) compounds, was discussed as main cause of the new type of forest damage (Schöpfer & Hradetzky 1984, Schütt et al. 1983, Ulrich 1984). Measures to mitigate air pollution as well as the establishment of the "International Co-operative Programme on Assessment and Monitoring of Air Pollution Effects on Forests" (ICP Forests) in 1985 were initiated under the Convention on Long-Range Transboundary Air pollution (CLRTAP) by the Economic Commission for Europe of the United Nations (UNECE). Methods for the forest condition survey are widely harmonised and standardised throughout Europe and are recorded in the ICP Forests manual (Eichhorn et al. 2016), which has continuously been subject to updates since its first publication in 1985. The survey is performed on the wide-scale monitoring plots (Level I), which were established wherever forest coincided with a $16 \times 16$ km grid over Europe (Ferretti et al. 2010). The forest condition survey is primarily based on defoliation, which denotes the loss of needles or leaves in the crown of a tree compared to a local or absolute reference tree with full foliage. Defoliation represents the most widely used indicator for the assessment of tree condition and vitality (Eichhorn et al. 2016, Ferretti 1997, Innes 1993). The estimation of defoliation takes place visually using binoculars. A quality assurance programme including e.g. national training courses (Eickenscheidt & Wellbrock 2014; Eichhorn et al. 2016) has been initiated in order to control consistency and reproducibility of defoliation data.

In Germany, the condition of forest trees was recorded first in 1984 and has been carried out annually throughout Germany since 1990. Grid densifications in addition to the $16 \times 16$ km grid are common within German federal states. In addition, changes of the grid over time occurred. Irregular grids and thus irregular time-series represent one of several characteristics of the defoliation data which make the spatio-temporal evaluation statistically challenging. Geoadditive models (Fahrmeier & Lang 2001, Kammann & Wand 2003) and generalized additive mixed models (GAMMs) (Lin & Zhang 1999) have been proposed for the evaluation of defoliation data. Inference can be based either on full Bayesian posterior analysis using Markov chain Monte Carlo (MCMC) techniques or on empirical Bayesian posterior analysis using mixed model techniques. Approaches were suggested for binary data (Fahrmeir & Lang 2001, Musio et al. 2007, Musio et al. 2008) and for ordinal data (Augustin et al. 2007, Kneib & Fahrmeir 2011). Up to date, spatio-temporal modelling of continuous defoliation data has only been published by Augustin et al. (2009) for Baden-Wuerttemberg. The authors used GAMMs and inference was based on mixed model methodology. These GAMMs are also promising regarding spatio-temporal modelling of defoliation throughout Germany. One of the most important aspects of spatio-temporal modelling is the handling of the spatio-temporal trend and the interaction of space and time. In the proposed model this aspect can be managed using the scale invariant tensor product, which is a three-dimensional smoothing function of space and time (Wood 2006a, Wood 2017). The tensor product allows modelling of data derived from irregular grids as well (Augustin et al. 2009). Models using the tensor product are most likely superior compared to models in which the spatial and temporal effects enter the model additively, such as models proposed by Kneib & Fahrmeir (2011), since it is unlikely that the spatial trend of defoliation is additive in time (Augustin et al. 2009). Another advantage of GAMMs is that (non-)linear effects of influencing parameters can also be modelled by using smoothing functions (Augustin et al. 2009, Wood 2017). In the case of defoliation it is well known that the level of defoliation depends on tree age (Klap et al. 2000, Seidling 2007). Thus, tree age can be considered in the model and trends solely based on tree age can be separated from trends caused by other parameters (e.g. air pollution, weather conditions). GAMMs further support a wide range of correlation structures (Wood 2017).

Annual descriptions of the time trend of defoliation, considering the 16 x 16 km grid only, are published by the Federal Ministry of Food and Agriculture (BMEL; e.g. BMEL 2017). Statistical evaluations of the spatio-temporal development of defoliation in Germany are hitherto lacking. Age-adjusted spatio-temporal trends are crucial for identifying regions with high mean defoliation. In addition, the suitability of the 16 x 16 km grid as well as of denser grid resolutions with respect to nationwide spatio-temporal modelling have not been examined for Germany up to date and generally methods and results of grid examinations regarding the forest condition survey have rarely been published (Köhl et al. 1994, Saborowski et al. 1997). Therefore, the present study aims to i) estimate the spatio-temporal trends of defoliation of the main tree species Norway spruce (*Picea abies* (L.) Karst), Scots pine (*Pinus sylvestris* L.), European beech (*Fagus sylvatica* L.) and pedunculated and sessile oak (*Quercus robur* L. and *Q. petraea* (Matt.) Liebl., treated together) in Germany from 1989 to 2015 using GAMMs and all available grid densities as well as to iii) examine the suitability of different monitoring grid resolutions, which have been used in the past, for spatio-temporal modelling using GAMMs by comparing the respective prediction errors.

## 2 Materials and methods

### 2.1 Data of the forest condition survey

Annual data of the forest condition survey of Germany is available from 1989 to 2015. Tree defoliation represents the main parameter of the survey and is given in 5% classes from 0% (no defoliation) to 100% (dead tree). Further parameters such as fructification and abiotic and biotic damage causes are also recorded. The investigated trees have to belong to Kraft class 1 (dominant) to 3 (subdominant), hence no suppressed trees are considered. The federal states are responsible for data collection, which is mandatory for the 16 x 16 km grid throughout

Germany. From 2006 to 2008 during the second national forest soil inventory (NFSI II), the forest condition survey was conducted nationwide on an 8 x 8 km grid with exception of the federal states Rhineland-Palatinate (4 x 12 km + 16 x 16 km), Saarland (2 x 4 km) and Schleswig-Holstein (8 x 4 km), which had denser grids. Data of the corresponding denser grids are further available for Baden-Wuerttemberg, Hesse, Lower Saxony and Saxony-Anhalt from 2005 to 2015, for Mecklenburg-Western Pomerania from 1991 to 2015, for Rhineland-Palatinate additionally from 2009 to 2010 and for Saarland from 2009 to 2015. A partial data set of the denser grid is further provided by Bavaria from 2009 to 2015. Two federal states changed their initial grid to coincide with the grid of the national forest inventory (Bavaria in 2006 and Brandenburg in 2009). In the following evaluations GAMMs were only used for the four main tree species. Norway spruce represented the most frequent tree species of Germany with 32.0% of all trees on the 16 x 16 km grid, followed by Scots pine with 29.1%, European beech with 16.2% and oak with 6.2%. The two oak species *Q. robur* and *Q. petraea* were regarded together due to occurrence of hybridization. It needs however to be kept in mind that 16% of the trees did not belong to the main tree species. Evaluations were carried out on plot level. The mean stand age for one tree species of one plot was estimated from the ages of single trees. In case where the youngest or oldest tree of a plot deviated more than 20 years from the plot mean, the stand age was classified as 'irregular'. For plots without single tree age specification (25% of the plots, e.g. systematically for Bavaria from 1989 to 2005), the non-species-specific stand age of the plot given in the database was used. Hence, for 98% of the plots with at least one tree belonging to the main tree species, a stand age (other than 'irregular') was available. The mean stand age (median weighted according to the number of trees per plot) regarding the 16 x 16 km grid differed among the species and this was primarily due to more very old trees in the deciduous species. In 2015 (1989-2015) the mean stand ages of spruce, pine, beech and oak amounted to 73 (70), 86 (70), 104 (90) and 110 (103) years, respectively. For all species the increase in age from 1989 to 2015 was notably lower than 26 years. Pine trees growing in the Northern Lowland were on average slightly younger than pine trees in the remainder of Germany. Each plot having at least one tree belonging to the main tree species and having a stand age was used for the statistical evaluations. Therefore, the total number of unique plots ranged from 290 in 1989 (only 16 x 16 km grid without former Eastern Germany) and 392 in 1990, respectively (only 16 x 16 km grid and entire Germany), to 1807 in 2008 (16 x 16 km grid and grid densifications in all federal states).

## 2.2 Spatio-temporal model for mean defoliation

We modelled the spatio-temporal defoliation monitoring data by species using a GAMM (Lin & Zhang 1999, Wood 2006a, Wood 2017). This spatio-temporal model was based on the one developed in Augustin et al. (2009):

$$logit\, E(y_{it}) = logit(\mu_{it}) = f_1(stand\ age_{it}) + f_2(easting_i, northing_i, year_t) \qquad (1)$$

where $y_{it}$ is the mean defoliation of one of Norway spruce, Scots pine, European beech or oak for sample plot $i = 1, \ldots, n$ and for year $t = 1, \ldots, 27$, averaged over all trees of the respective species at sample plot $i$. Before averaging, the defoliation class of a single tree was converted into a continuous variable by using the midpoint of the class. The logit link was used since defoliation represents an estimated percentage and the logit link ensured that fitted values were bounded in (0,1). The function $f_1$ is a one-dimensional smooth function of stand age$_{it}$ using a penalized cubic regression spline basis. The function $f_2$ is a three-dimensional smooth function of the year and of the coordinates (easting and northing of the Gauß-Krüger coordinate system, GK4), which is a tensor product smoother constructed from a two-dimensional marginal smooth for space and a marginal smooth for time (Augustin et al. 2009). The marginal bases are a two-dimensional thin plate regression spline basis for easting and northing and a cubic regression spline basis for year. The tensor product of the two marginal smooths is used so that different penalties for space (meter) and time (years) are used (Wood 2006a, Wood 2017). For the error term ϵ a multivariate Normal distribution was assumed as N(**0**, $\sigma^2$ **Λ**). The covariance matrix **Λ** is a block diagonal matrix with the i[th] subvector $\epsilon_i$ having covariance matrix $\mathbf{\Lambda}_i$, which is related to the residuals of one plot *i* over time. The covariance matrix $\mathbf{\Lambda}_i$ further contains on the diagonal the weights $1/\alpha_{it}$, where $\alpha_{it}$ is the number of trees assessed at plot *i* and year *t*. The temporal correlation was modelled by a first order autoregressive-moving average process (ARMA(1,1)) (Pinheiro & Bates 2000), i.e. $\epsilon_{it} = \varphi\epsilon_{it-1} + \mathrm{p}c_{it-1} + c_{it}$ where φ and p represented correlation parameters and $c_{it}$ follows a Normal distribution with an expected value of zero. The ARMA(1,1) process was most suitable for all four tree species according to model selection.

Parameter estimation can be carried out as for a GLMM using penalized quasi-likelihood with standard mixed modelling software (Augustin 2009, Lin & Zhang 1999, Wood 2004, Wood 2017). This is possible since the GAMM in equation (1) corresponds to a generalized linear mixed model (GLMM) because the smooth functions $f_1$ and $f_2$ can be rewritten as:

$$logit(E(y_{it})) = \mathbf{X}_{it}\mathbf{\ss} + \mathbf{Z}_{it}\mathbf{b} \qquad (2)$$

The matrices **X** and **Z** are the design matrices containing the basis functions evaluated for each observation at plot *i* and year *t*. The matrix **X** contains the parts of the basis function to which the unpenalized coefficients **ß** apply, e.g. for $f_1$ this is a straight line and the matrix **Z** contains all parts of the basis functions to which the penalized coefficients **b** apply. The vector **b** is a vector of random effects following a Normal distribution with mean zero and an unknown positive-definite covariance matrix ψ.

For variance and trend estimation we used the Bayesian representation of the GLMM in equation (2). This is done by interpreting the choice of smoother and penalty as making prior assumptions about the smoothness of the true function. So the penalties can be expressed as prior distributions on the functions $f_1$ and $f_2$. Then by using Bayes theorem, a posterior distribution of the model parameters is obtained (Augustin et al. 2009, Silverman 1985, Wahba 1983, Wood 2006b, Wood 2017) and this is a multi-variate Normal distribution. Hence by sampling from the multi-variate Normal posterior distribution of the model parameters the predictive distributions of any quantity of interest can be obtained; the lower and upper 95% quantiles constitute the credible interval for the quantity of interest. We use these type of intervals because in the context of GAMMs the Bayesian credible intervals have been shown to have good coverage properties (Silverman 1985). All evaluations were performed using R 3.2.2 (R Development Core Team 2015). For the spatio-temporal modelling of defoliation the R package mgcv (Version 1.8 10; Wood 2015) was utilized. In the appendix we give example R code for the described analysis.

Residual correlation was examined by using diagnostic plots in which normalized residuals were investigated in space and time (see Augustin et al. 2009). This included empirical semi-variograms (R package geoR by Ribeiro & Diggle (2015)) as well as empirical (partial) autocorrelation functions (R package nlme by Pinheiro et al. (2015)). Although stand age is confounded with space and time, collinearity between the parameters stand age and year did not occur as the correlation between stand age and year was low. The Bayesian information criterion (BIC) was used for model comparison and selection (Schwarz 1978) due to its suitability for data having a high sample number compared to the number of model parameters. Further details on the statistical methodology are given in Augustin et al. (2009).

## 2.3 Trend estimation

As described above we generated predicted values of mean defoliation $\hat{\mu}_{itp} = \hat{y}_{itp}$ from the posterior predictive distribution of $y_{it}$ and averaging over $i$ then yields the mean defoliation in year $t$

$$\hat{\mu}_{tp} = \hat{y}_{tp} = \frac{1}{n}\sum_{i=1}^{n}(\hat{y}_{itp}) \qquad (3)$$

for the $p^{th}$ draw from the posterior predictive distribution. The median and the 2.5% and 97.5% quantiles were calculated for the spatial ($\hat{\mu}_{itp}$) and temporal trend ($\hat{\mu}_{tp}$). For the trend estimation plots were not weighted by number of trees per plot because on the one hand weighting would down-weight plots without pure stands, which would be an undesirable characteristic of the estimator (Augustin et al. 2009), and on the other hand we were interested in a statement for the entire area of Germany. For parameter estimation of the spatio-temporal model shown in equation (1) and (2) available data of all grid resolutions were used. For time trend estimation we generated a predictive distribution as described above using an area- and species-representative grid. We used the 16 x 16 km grid of the forest condition survey i) of 2015 (grid 1) and ii) of the corresponding year (grid 2). In the first case (grid 1), the grid of 2015 was transferred to all other years (1989-2014), which offers the possibility to exclude the age and grid effect. In the second case (grid 2), the actual observed stand age of each plot was considered as well as changes in the 16 x 16 km grid (shift, elimination/addition of plots).

## 2.4 Simulation study for the grid examination

In Germany, the federal states are obliged to provide the forest condition data of the 16 x 16 km grid for the annual nationwide time trend evaluation by the BMEL. However, it is yet unclear if this grid is sufficient for nationwide spatio-temporal trend modelling of defoliation. Therefore, the mean prediction error (MPE) of the defoliation estimates was estimated i) for the 16 x 16 km grid and the 8 x 8 km grid (the federal states having denser grids than 8 x 8 km (Rhineland-Palatinate, Saarland and Schleswig-Holstein) were not included for MPE estimations for the 16 x 16 km grid and the 8 x 8 km grid) (approach I) and ii) for the 16 x 16 km grid and for all grid resolutions available, which corresponded to the highest available grid density (all federal states were included for MPE estimations for the 16 x 16 km grid and for all grid resolutions available) (approach II). The parameters of the spatio-temporal model described in equation (1) and (2) were estimated using all available data of all federal states. For the simulation study the estimates of the model fitted to the observed defoliation data were assumed to be the truth. As described above (section 2.2 and 2.3) data were simulated from the multivariate Normal posterior distribution of these parameters. We draw $s = 1, \ldots, nsim$ ($nsim = 40$) set of parameters and the $s^{th}$ draw yields a realisation of the predictive distribution of the response for all possible grid points and all years:

$$\hat{y}_{its} = logit^{-1}\left(\boldsymbol{X}_{it}\widehat{\boldsymbol{\beta}}_s + \boldsymbol{Z}_{it}\widehat{\boldsymbol{b}}_s\right) = logit^{-1}(\hat{\eta}_{its}) \tag{4}$$

where $\hat{\eta}_{its}$ is the linear predictor simulated from the predictive distribution and thus the simulated data value of defoliation $y_{its}$ at plot $i$ and year $t$ is:

$$y_{its} = \frac{\exp(\hat{\eta}_{its})}{1 + \exp(\hat{\eta}_{its})} + \varepsilon_{its} \tag{5}$$

For the generation of the error term, the parameter estimates relating to the ARMA error model $\varepsilon_{its} \sim N(\boldsymbol{0}, \hat{\sigma}^2 \widehat{\boldsymbol{\Lambda}}_i)$ were used. Moreover, the weights $1/\alpha_{it}$ were included. Under this simulation scheme, defoliation values lower than 0 and higher than 1 were possible and were truncated to 0 and 1, respectively. Exploratory plots showed that the simulated defoliation data reflected the actual observed data well. We simulated data for all sample plots available in 2008 (year with most plots) (for approach I the federal states Rhineland-Palatinate,

Saarland and Schleswig-Holstein were excluded). The stand age of each plot was set to the stand age of 2008 since the mean age stayed relatively constant over the whole period from 1989 to 2015. In the next step, a survey sample was taken from the simulated data and the GAMM was fitted to the survey sample data. The sample taken was either the sample obtained from the 16 x 16 km grid or from the 8 x 8 km grid of approach I or from the 16 x 16 km grid or from the denser grid of approach II. Subsequently, the time trend $\hat{y}_{ts}$ for Germany was estimated from the simulated sample data using equation (3). This was then compared with the assumed true $\mu_t$, as estimated from the original data in equation (3), to estimate the MPE per year *t*:

$$MPE_t = \frac{1}{nsim} \sum_{s=1}^{nsim} (\hat{y}_{ts} - \mu_t)^2 \qquad (6)$$

Finally, the MPE per plot *i* and year *t* was estimated by comparing the estimated mean defoliation $\hat{y}_{its}$ with the assumed true $\mu_{it}$ (from equation (1) where the parameters were replaced by their estimates from the model fitted to the actual data):

$$MPE_{it} = \frac{1}{nsim} \sum_{s=1}^{nsim} (\hat{y}_{its} - \mu_{it})^2 \qquad (7)$$

In the following the square root of the MPE is used (RMPE). The $RMPE_{ti} > 5\%$ was considered for the years 2006 to 2015 since we focused on the suitability of the present grid.

# 3 Results

## 3.1 Age-adjusted spatio-temporal trend of defoliation

The best fit of the spatio-temporal model was obtained for spruce (adjusted $R^2$ of 0.54, n = 10182), followed by beech (adjusted $R^2$ of 0.47, n = 9283), oak (adjusted $R^2$ of 0.47, n = 6098) and pine (adjusted $R^2$ of 0.41, n = 9252) (Table 1). Both, the stand age and the space-time component, showed a highly significant effect on defoliation ($p < 0.0001$) (Table 1). For comparison, adjusted $R^2$ of models not including the stand age and space-time component ranged between 0.009 (spruce) and 0.031 (beech). The effect of stand age on defoliation differed among the tree species (Fig. 1). A nearly linear increase of defoliation with increasing stand age was observed for spruce and beech. For pine, defoliation increased until a stand age of approximately 40 years and hardly any dependency did occur for older pine trees. For oak, likewise as for pine, defoliation clearly increased until a stand age of approximately 60 years whereas defoliation only slightly increased with further increasing stand age.

Defoliation of spruce remained more or less the same from 1989 to 2015 (Fig. 2). The mean defoliation was 20.2% (grid 1) and 18.9% (grid 2), respectively. Highest mean defoliation was observed in 1992 and in the two following years as well as after 2003 (grid 1: 23.3% in 2004). Lowest defoliation occurred in 1989 (grid 1: 17.9%). At the beginning of the observations, the defoliation estimated for the grid and median age of 2015 (grid 1) was higher than the defoliation estimated for the actual observed grid and age (grid 2) due to differences in the grids and especially due to a lower actual stand age at the beginning of the time series. Over time both time trends overlapped. According to the association of defoliation in spruce and stand age (Fig. 1), defoliation of 150 years old spruce was significantly higher than of 50 years old spruce (Fig. 2). The course of the time trend was the same for both ages but it was more pronounced in the older spruce trees. The credible interval for the time trend of the 150 years old trees was wider than that of younger trees because trees of 150 years are less common in spruce.

Pine showed the lowest mean defoliation regarding the four main tree species with 18.2% (grid 1; 16.6% for grid 2). Highest defoliation was observed at the beginning of the 1990ies (1991: 25.1% for grid 1) (Fig. 2). Defoliation decreased until the mid 1990ies and stayed more or less unchanged up to 2015. A slight increase was observed in the years following 2003. Due to the weak association of defoliation in pine with stand age compared to the other species (Fig. 1), differences between the time trend of 150 years old and of 50 years old pine were low compared to the other tree species (Fig. 2).

Defoliation of beech was higher than those of the two coniferous tree species (grid 1: 24.6%, grid 2: 21.3%). Defoliation on average showed no clear trend but pronounced fluctuations were observed (Fig. 2). Peaks of defoliation occurred in 1992, 2000, 2004-2006, 2010-2011, 2014) with highest defoliation in the years 2004 and 2005 (grid 1: both years 29.1%). Other than for spruce, pine and oak, highest defoliation of beech was observed in years with high fructification, so called mast years. Beech trees of 150 years had notably higher defoliation than beech trees of 50 years (Fig. 2).

Oak showed the highest mean defoliation of all species with 26.9% (grid 1; 23.0% for grid 2) (Fig. 2). Defoliation of oak was lowest in 1989 (grid 1: 19.3%) and subsequently increased until 1993 (28.8%). In the following years defoliation remained on a high level with a peak in 2004 (30.5%) but a slight decrease could be observed in 2002 and in the last years (Fig. 2). The credible intervals were wider as compared to the other main species since oak is less frequent. Differences between the time trend of 50 years old and 150 years old oak trees were evident but slightly lower than for spruce and beech (Fig. 2).

Defoliation varied spatially in Germany between 1989 and 2015 (Fig. 3-6). For spruce, defoliation was high in north-eastern Germany at the beginning of the 1990ies (Fig. 3). It was further slightly higher in south-eastern Germany in 1992. From 1994 to 2002 defoliation of spruce was comparably low everywhere in Germany. In 2003 defoliation slightly increased in southern Germany. Beginning in 2005 highest defoliation was found in south-western Germany with highest defoliation in 2006 and 2007 in particular in south-western Baden-Wuerttemberg (e.g. Black Forest). Between 2005 and 2015 defoliation was generally lower in the eastern part of Germany (Mecklenburg-West Pomerania to Bavaria) than in the western part (Schleswig-Holstein to Baden-Wuerttemberg).

For pine, high defoliation >25% occurred in north-eastern Germany as well as in parts of the border region of southern Germany at the beginning of the 1990ies (Fig. 4). In the mid 1990ies a difference between the Northern Lowland with low defoliation and the remainder of Germany with slightly higher defoliation showed up and remained until 2015. Regarding the remainder of Germany, again south-western Germany appeared as region with highest defoliation beginning in 2003.

For beech, at the beginning of the time series high defoliation was found in north-eastern Germany (Fig. 5). In 1992 defoliation > 25% occurred in central and southern Germany and in 2000 in north-western Germany. After 2003 high defoliation was observed in large areas of central and southern Germany. South-western Germany again emerged as region of partly very high defoliation but comparably high defoliation also extended to Hesse and up to the south of Lower Saxony.

For oak, at the beginning of the 1990ies defoliation was also highest in north-eastern Germany but it was also high in south-eastern Germany (Fig. 6). Beginning in the mid 1990ies, besides south-western Germany, the lowland north of the low mountain ranges (especially in North Rhine-Westphalia) emerged as area of recurring high defoliation.

Summing up, a shift of high defoliation was demonstrated for all species with high defoliation in north-eastern Germany at the beginning of the time series and high defoliation in south-western Germany as from 2003.

**3.2 Grid examination**

The $RMPE_t$ of the 16 x 16 km grid was lowest for pine with 0.9% (approach I and II) followed by spruce with 1.0% (approach I and II), by beech with 1.6% (approach I) and 1.3% (approach II), respectively, and was highest for oak with 1.9% (approach I) and 1.8% (approach II), respectively (Table 1 and Table 2). The $RMPE_t$ of the 8 x 8 km grid (approach I) and of the grid densification (approach II) were lower in all cases and ranged from 0.5% to 1.1% (Table 1 and Table 2). Through the years no trend was observed for the $RMPE_t$ of the 16 x 16 km grid whereas the $RMPE_t$ of the denser grids slightly decreased from 1989 to 2015 for all tree species. Although the mean $RMPE_t$ as well as the mean $RMPE_{it}$ (Table 1 and Table 2) was low for the four main tree species and all grids (16 x 16 km grid and 8 x 8 km grid/grid densification), single sample plots or regions showed $RMPE_{it} > 5\%$ (Table 1 and Table 2). No or only two plots with a $RMPE_{it} > 5\%$ were found for spruce, pine and beech for the 8 x 8 km grid and the grid densification, respectively (Table 1 and Table 2). Moreover only few oak plots had $RMPE_{it} > 5\%$ regarding these densified grids. However, regarding the 16 x 16 km grid in particular many oak plots had $RMPE_{it} > 5\%$ (Table 1 and Table 2). Spruce showed the lowest number of plots having $RMPE_{it} > 5\%$. Regional differences among tree species occurred regarding the $RMPE_{it}$ of the 16 x 16 km. For spruce highest uncertainties existed in the Eifel region, in the Alpine region and in the north of the Black Forest (Fig. 7). These uncertainties did not occur if denser grids were used. For pine different regions showing uncertainties were found regarding the 16 x 16 km grid. Weak points were particularly the southern and also northern parts of Germany (Fig. 8). For beech high uncertainties did especially occur at the borders of Germany (northern and north-eastern Germany, Saarland, south-western Baden-Wuerttemberg, the Alpine region and the Bavarian Forest) (Fig. 9). In case of the denser grids, high uncertainties did not exist for pine and beech. For oak high uncertainties were found scattered over almost entire Germany regarding the 16 x 16 km grid (Fig. 10). The $RMPE_{it}$ of oak plots frequently amounted to even more than 10%. In case of the denser grids, only single plots showed $RMPE_{it} > 5\%$ as well as the region of and around the Bavarian Forest. In conclusion, the 16 x 16 km grid was sufficient for the time trend calculation for the four main tree species, although a higher uncertainty was observed for oak. The most frequent species spruce and pine showed the lowest prediction error and the least frequent main tree species oak showed the highest error. For spatio-temporal trends deficiencies occurred for all species regarding the 16 x 16 km grid.

# 4 Discussion

### 4.1 Spatio-temporal modelling of defoliation

GAMMs proved to be a good choice for spatio-temporal modelling of defoliation data. Statistically sound estimates of mean defoliation in space and time with appropriate credible intervals could be produced. Use of a three-dimensional smoothing function of space and time allowed for high flexibility regarding changes in the grid and thus, defoliation data of all available grid densifications could be included. Our study revealed that stand age explained up to half of the observed variability in defoliation. Adjustment for stand age was possible using GAMMs and carried out in order to identify hot spots of high defoliation not merely resulting from the age effect. The age effect primarily determined the general level of defoliation whereas fluctuations in the time series of defoliation were most likely associated with weather conditions.

### 4.1.1 Age effect: determinant of the defoliation level

A pronounced effect of age on defoliation was common to all investigated tree species. This effect has frequently been mentioned in the literature in particular for spruce (Eichhorn et al. 2005, Riek & Wolff 1999, Seidling & Mues 2005, Zirlewagen et al. 2007). The species-specific age effects mostly corroborated results published by Augustin et al. (2009), Klap et al. (2000) and Seidling (2001). The species-specific association between defoliation and age might be primarily attributed to the demand of light, the stand situation, the stand development and the forest management (thinning, felling under mature canopy) typically found for the corresponding species. Beech is known to be shade-tolerant and the almost linear increase in defoliation with age might reflect the comparably slow but continuous vertical growth of beech (Pretzsch et al. 2015). Beech crowns shadowed by neighbouring trees or beech trees with lower social status were shown to have lower defoliation (Seidling 2004). This finding is in line with observations from the federal states Rhineland-Palatinate and Saarland where old trees growing in canopy-closed stands were found to have lower defoliation than old trees growing in cleared stands (MULEWF 2015). Additionally, old beech trees usually were found in cleared stands with natural rejuvenation in Rhineland-Palatinate (MULEWF 2015), which is also the case for other federal states. The higher defoliation of these old beech trees might be the result of being more exposed to wind, being more susceptible to drought stress due to higher water consumption and especially of having more intensive fructification than beech trees in canopy-closed stands. Oak and pine belong to the light-demanding tree species. Unlike for beech, Seidling (2004) reported higher defoliation for oak and pine trees with lower social status. For both tree species a similar asymptotic association between defoliation and age was found. Light-demanding tree species show a notable vertical growth in young stands whereas the growth rate decreases with increasing age (Pretzsch et al. 2015). Pine trees grow faster than oak trees but the time point of decrease of growth rate is also reached earlier. Assessed pine trees mainly grew in pure stands (in particular in the Northern Lowland) whereas oak trees were mainly found in mixed stands. In Rhineland-Palatinate, old trees of both species were mainly found in canopy-closed stands (MULEWF 2015), which is also true for other federal states like Baden-Wuerttemberg. Spruce belongs to the semi-shade tree species and the age effect is similar to the curve progression of beech. Very old trees of spruce were often found at sites having extreme climatic conditions or being very poor, e.g. spruce trees growing in the Bavarian Alps. Hence, confounding between age and site conditions cannot be excluded and may partly be responsible for the linear increase of defoliation observed for very old trees of this species. The estimated effects of age on defoliation represent associations as they were found on average for the tree species. However, old trees and plots having several old trees of one species for which low defoliation was observed in the long term could be found for each tree species. Natural senescence plays a role for the observed age effect. Changes in crown morphology, i.a. because of sexual maturity and accompanied recurring fructification, represent one aspect of natural senescence. However, trees do probably not age in close relation to time but the social status and stress factors determine their senescence (de Vries et al. 2014, Pretsch & Rais 2016). Several authors proposed an accumulating effect of multiple stress factors with age resulting in an enhanced sensitivity of older than of younger trees (Klap et al. 2000, Seidling & Mues 2005, Solberg 1999). This assumption underlines that occurrence of healthy old trees is possible as well as that old trees can regenerate under favourable conditions (MULEWF 2015). However, the selective vulnerability to a certain stress factor does presumably not increase in older trees compared to younger trees (Klap et al. 2000).

In a Europe-wide evaluation of defoliation, significant effects of age on defoliation were found for all four main tree species only in Germany and France, whereas no age effect was observed for these tree species in several other European countries (Seidling & Mues 2005). The observation was corroborated by Vitale et al. (2014) for spruce. First, a possible reason for this observation might be that countries like Germany and France have a large range of ages and a sufficient high sample size for each age (class), which together build the basis for exploration of existing relationships. Second, defoliation of senescent tree crowns might be assessed differently since senescence might be considered in the reference system of some countries (Klap et al. 2000, Seidling & Mues 2005). Finally, the age effect may be an apparent effect which reflects the stand structure, management practices and the mean long-term stress level found in the respective country. Presumably all three reasons contribute to the explanation of the observed differences among countries.

In the present study it was shown that the general level of defoliation was primarily determined by the mean age and the species-specific association between defoliation and age. Hence, pine had the lowest and oak the highest mean defoliation. The age adjustment for time trend estimation using a stand age of 50 years and 150 years, respectively, underlined that the differences in the mean level of defoliation among the tree species was primarily a result of the age effect. Oak, beech and spruce trees of 150 years on average showed the same level of defoliation whereas old pine trees had a notably lower level of defoliation, which could be attributed to the species-specific association with age. Oak trees having a stand age of 50 years, however, exhibit a higher level of defoliation compared to spruce, pine and beech trees, which in turn had similar defoliation. Comparison of the actually existing time trend that arose using the true stand age of trees on the 16 x 16 km grid that was monitored in the respective year (grid 2) with the age- and grid-adjusted time trend (grid 1) further corroborated the effect of age on the defoliation level. Differences in the time course between both time series mainly derived due to different grids whereas differences in the level of defoliation could be attributed to the differences in stand age. Both effects were strongest at the beginning of the time series since grids deviated most from the grid in 2015 (e.g. shift of grids) and trees on average were younger and thus lower defoliated than in 2015. Thus, the present study underlines that an age adjustment is necessary for time trend as well as for spatio-temporal trend estimation in order to obtain time trends which do not mirror increasing defoliation as consequence of aging and to identify regions with longer-term high defoliation that could not primarily be ascribed to the age effect.

**4.1.2 Weather conditions: drivers of spatio-temporal defoliation trends**

In summer 2003 an extreme heat wave and drought occurred in large parts of Europe, which had a strong impact on forest ecosystems (Allen et al. 2010, Lindner et al. 2010). The severe drought stress in 2003 climaxed after the forest condition survey. Notably increased mean defoliation was observed in 2004 for all investigated tree species. For spruce, beech and oak, highest mean defoliation was even found in this year. In 2004, highest defoliation occurred in south-western Germany regarding all species and this region continued to be the area of highest defoliation until the end of the monitored period. Though entire Germany was affected by drought stress in 2003, the strongest water deficits compared to the long-term reference period (1961-1990) were found in south-western Germany and particularly in the area of the Black Forest (Anders et al. 2004, Eickenscheidt et al. submitted). This region was further on characterized by the most distinct drought events in the following years compared to the remainder of Germany (Eickenscheidt et al. submitted). Investigations regarding associations between influencing parameters and defoliation were simultaneously conducted in another study by us using the same data and GAMMs (Eickenscheidt et al. submitted). The study revealed strong statistical associations between weather conditions and defoliation of all four tree species. An increase in defoliation apparently occurred for all species at positive temperature deviations from the long-term mean (1961-1990) of more than 1°C particularly in combination with negative precipitation deviations. For the German 16 x 16 km grid, associations between defoliation and deviations from long-term means of temperature and precipitation have already been published previously considering the years 1990 to 2004 (Seidling 2007). Moreover, a significant influence of climatic factors and an increase in defoliation with drought in Europe have been reported, e.g. for France by Ferretti et al. (2014), for Spain by Carnicer et al. (2011), for Switzerland by Zierl (2004) and Europe-wide by Klap et al. (2000). Lagged effects, especially drought of the previous year, and cumulated drought of several preceding years show a major influence on defoliation in the following year (Ferretti et al. 2014, Klap et al. 2000, Seidling 2007, Zierl 2004), which was corroborated by our findings.

In general, defoliation of the investigated four tree species developed differently in time and space over the monitored period. Differences in the geographical distribution, in drought tolerance as well as in further species-specific influencing factors might be reasons for this observation. Norway spruce represents the most common tree species of Germany, which predominantly grows in moist and cooler regions of the low mountain ranges and the Alpine foreland but which is rarely found in the Northern Lowland. The sensitivity of spruce to water stress is commonly known and mostly ascribed to the shallow root system (BLAG-FGR 2014, Ellenberg 1996). This tree species showed comparably low mean defoliation with low temporal changes between 1989 and 2015 as well as low spatial variation within Germany. The highest mean defoliation occurred in 2004, but defoliation remained elevated in 2005, 2006 and 2007. In 2009 the defoliation level prior to the drought event was reached again. Jonard et al. (2012) also reported an increase in defoliation of spruce in the Ardennes until 2009 subsequently to the drought in 2003. In general, the needle loss is still visible years after the event because spruce trees keep up several needle sets. Cumulated drought was demonstrated to be most important for spruce trees in Switzerland (Zierl 2004), which could be corroborated by our results.

Scots pine is the main tree species growing in the Northern Lowland but it can also be found in the remainder of Germany. This tree species in general showed the lowest mean defoliation and hardly any temporal changes in defoliation were observed between the mid 1990ies and 2015. However, since the mid 1990ies, spatial differences between the Northern Lowland with low defoliation and the remainder of Germany with higher defoliation were prominent. The nationwide mean defoliation was only slightly higher between 2004 and 2006. This increase was solely obvious in central to southern Germany. Pine trees are generally known to be relatively drought-tolerant (Ellenberg 1996), which can be ascribed to several mechanisms like a deep taprooting system and early and rapid stomata closure (e.g. Seidling 2007 and references therein). However, notable temperature surplus and precipitation deficits as observed in 2003 particularly in southern Germany probably caused visible drought stress even in pine trees. The high defoliation which was observed in parts of north-eastern Germany (Mecklenburg-West Pomerania, Brandenburg and Saxony-Anhalt) particularly in 1991 was probably caused by severe insect infestation (Eickenscheidt et al. submitted). Seidling 2001 and Seidling & Mues 2005 also demonstrated the importance of insect infestations on defoliation of pine in Germany. In north-eastern Germany pure pine stands are common, which probably further promoted insect infestation. Methodological differences in defoliation assessment after the introduction of the forest condition survey in former Eastern Germany especially in Mecklenburg-West Pomerania however cannot be ruled out as reason for particular high defoliation (Riek & Wolff 1999).

European beech is the most common deciduous tree species in Germany and similar to spruce is mainly distributed in the moist mountainous regions and comparably rarely represented in the Northern Lowland. Sensitivity of beech to drought is well known, however drought resistance varies among beech populations (Bolte et al. 2016). Beech usually develops from natural rejuvenation and thus is adapted to the site conditions (BLAG-FGR 2014). Mean defoliation of beech was higher than those of the coniferous species and showed clear fluctuations, which were primarily coincided by occurrence of common to abundant fructification of beech trees. The years 1992, 2000, 2004, 2006, 2009, 2011 and 2014 represented pronounced mast years for beech and also showed high mean defoliation. Findings were in line with observations by e.g. Eichhorn et al. (2005) and Seidling (2007). Weather conditions in the previous early summer determine the production of flower buds and leaf buds, respectively. Hence, fructification is closely linked to higher defoliation. Furthermore, small leaves are common due to the high demand of nutrients for fructification (MULEWF 2011). Mast years are frequently observed after a warm and dry summer in the previous year, which had repeatedly been fulfilled during the investigated time period. A literature review by Paar et al. (2011) further demonstrated an increase in mast years since 1988. Regional differences in annual fructification intensity were found. For example in 2000 common to abundant fructification occurred in western and north-western Germany, where highest temperatures combined with lowest precipitation was found in the previous year, and was accompanied by high defoliation in this part of Germany (Eickenscheidt et al. submitted). Highest mean defoliation regarding entire Germany was observed in 2004 and 2005 after the drought. Not before 2008 defoliation again reached the level of defoliation immediately before the drought, although annual leave fall in autumn eliminates direct carry-over effects from year to year unlike in coniferous trees. The combination of drought stress particularly in 2003 but also to a lesser extent in 2006 and substantial mast behaviour in 2004 and 2006 obviously required considerable time for regeneration of beech.

Pedunculate and sessile oak together represent the less common of the main tree species. Oaks grow from the Northern Lowland to the low mountain ranges but rarely in pronounced mountainous and cooler regions. Oak tolerates a wide range of climatic conditions and soil water availability. It can be found on soils with stagnant soil water but it is also known to be drought-tolerant due to its taproot system and fast stomatic response. Oaks showed the highest mean defoliation regarding the main tree species. In contrast to the other tree species, a recurrent pattern of defoliation possibly occurred. Lowest defoliation was observed at the beginning of the time series, in 2002 and in 2014 and 2015. Hence, starting with relatively low defoliation, defoliation increased, remained on a high level for some years, decreased and reached the low level again after 12 years before the same course started once more. Insect infestation was reported to be strongly associated with defoliation of oak trees (Eichhorn et al. 2005, Eickenscheidt et al. submitted, Seidling & Mues 2005). In how far development cycles of insects may be important for the observed repeated pattern needs further investigation. High defoliation was recurrently observed in the lowland adjacent to the low mountain ranges in particular in North Rhine-Westphalia (Westphalian Lowland). This area was also regularly affected by insect infestation. Highest defoliation occurred in 2004 after the drought. In this year insect infestation was widely observed in Germany and in particular in the lowland north of the low mountain ranges. Other than for beech and the coniferous tree species, defoliation significantly increased only in the year after the drought.

In conclusion, although the four tree species responded differently in time and space, drought stress seemed to be an apparent trigger of defoliation of all four species. It is further likely that weather conditions not only influenced fluctuations of defoliation directly (e.g. drought stress) but also indirectly by controlling fructification and propagation of insects. According to IPCC (2014), 'the period from 1983 to 2012 was likely the warmest 30-year period of the last 1400 years in the Northern Hemisphere'. In Central Europe, a further increase in the frequency of summer drought (high temperature combined with low precipitation), in precipitation in winter and spring, in early and late frost, in wind storms and in hail is expected as result of global warming (Lindner et al. 2010). An adaption of forest stands to the long-term average of the local water balance is generally supposed (Zierl 2004 and literature therein). However, forest trees show a particular sensitivity to climatic changes since their long life-span does not allow for rapid adaption (Lindner et al. 2010).

**4.2 Model-based approach for grid examination**

In Germany, annual nationwide time trend evaluation of defoliation by the BMEL is based on the data of the 16 x 16 km grid, which is part of the Europe-wide Level I grid. The federal states are obliged to deliver data of this grid. Our study revealed that the 16 x 16 km grid is in general sufficient to make a nationwide statement on time trends of defoliation for the four main tree species. A higher uncertainty exists for oak mainly due to being the less common main tree species. Spatio-temporal modelling is additionally necessary to identify hot spots of high mean defoliation at an early stage. For nationwide spatio-temporal trend estimation of defoliation using GAMMs, grids denser than 16 x 16 km (at least 8 x 8 km) are required for some regions for spruce, pine and beech and for entire Germany for oak. Mostly since the start of the NFSI II, several federal states have already provided defoliation data of their denser grids for the nationwide evaluations. In the present study spatio-temporal modelling was based on data of all available grids. It needs to be considered that statements concerning the prediction error of defoliation trends can only be made regarding the four main tree species as other tree species were not investigated.

Although grid examinations were carried out by federal states, results and methods were rarely published. For Lower Saxony it was shown by means of sample error and 90% confidence intervals that a 4 x 4 km grid is sufficient for this federal state (Saborowski et al. 1998). For Baden-Wuerttemberg a simulation study in 2006 revealed that the 16 x 16 km grid was not sufficient for spatio-temporal trend estimations (N. H. Augustin, personal communication). An 8 x 8 km grid including the 16 x 16 km grid points as fixed points whereas the remaining points would be alternating subsets of the 4 x 4 km grid was suggested. For Switzerland, Köhl et al. (1994) estimated sampling errors and reported that the loss of precision remained acceptable using an 8 x 8 km grid for nationwide evaluations whereas a strong deterioration was observed when using a 12 x12 km and 16 x 16 km grid, respectively. Design-based approaches (e.g. Köhl et al. 1994, Saborowski et al. 1998) as well as model-based approaches (N. H. Augustin (personal communication), Riek & Wolff 1997) were used for grid examinations. In the design-based approach the observed values are regarded as a sample from a fixed population, that is the observed values are regarded as fixed and the sample locations are regarded as the random quantity. Estimates and inference is based on a probability sample and is obtained from the resulting sampling distribution. The approach requires data to be derived from random sampling and to be independent from each other (Brus & Gruijter 1997, Lark & Cullis 2004). In contrast, in the model-based approach the population is regarded as random and the sampling locations are fixed (Brus & Gruijter 1997). Thus, random sampling and independence is not required but instead a spatial dependence needs to be modelled if present. Estimates and inference are based on the model and the model should mimic the data generating process. Data of the forest condition survey are obtained by systematic sampling on grids, which is a random sampling design and both the design-based and model-based approaches can be used for grid examination. Advantages of the model-based approach are that we can estimate the effects of explanatory variables on defoliation, we obtain spatio-temporal point estimates and we can predict, which makes it possible to estimate bias and prediction error, respectively, by means of simulation. Therefore, we applied the model-based approach using GAMMs. Since the simulation of the data and the prediction on basis of the simulated data were based on the same model (which is believed to be the true model), the prediction error estimated by us would rather underestimate the true prediction error than overestimate it.

# 5 Conclusions

In the present study generalized additive mixed models (GAMMs) were used for spatio-temporal modelling of defoliation data of the four main tree species of Germany from 1989 to 2015 as well as for examination of the suitability of the 16 x 16 km grid, which represents the basis grid of Germany. This is the first time the spatio-temporal modelling for entire Germany has been carried out although data has been collected since 1989. The model-based approach for grid examination turned out to be well appropriate for the given data and sample design. Weak points occurred for all four tree species regarding spatio-temporal trend estimation based on the standard 16 x 16 km grid. For oak, we recommend using at least an 8 x 8 km grid. However, the standard grid was generally sufficient for nationwide time trend estimation. GAMMs proved to be a statistically sound and highly flexible choice for spatio-temporal modelling of defoliation data. Stand age explained up to half of the observed variability in defoliation. The association between defoliation and age was species-specific but defoliation in principal increased with increasing age. Thus, the mean age and the species-specific relationship were mostly responsible for the general level of defoliation of the four tree species. The age effect was attributed to natural senescence, accumulation of stress with age and natural stand development as well as forest management. However, further investigations are necessary in order to understand what is behind the age effect and how it should be handled for evaluations of defoliation data. Adjustment for stand age was carried out in order to identify hot spots of high defoliation not merely resulting from the age effect. Spatio-temporal fluctuations were primarily the result of weather conditions. The present study demonstrated that the focus is moved from air pollution to the consequences of climate change. Exceptional drought stress occurred in 2003 and was associated with a notable increase in mean defoliation in the following years. Intensity and duration of the increase in defoliation were species-specific. Besides direct effects of weather conditions, indirect effects like increased fructification and mass propagation of insects are likely to play a major role for defoliation. The latter was especially of importance for the relatively drought-tolerant tree species pine and oak, whereas the combination of drought stress and substantial mast behavior probably led to considerable time for regeneration of beech. In recent years, south-western Germany turned out to be the region of highest mean defoliation regarding the four tree species. This region was affected by the strongest water deficit in 2003 compared to the reference period (1961-1990). Future measures should aim for a further reduction of soil acidification and mitigation of air pollution as well as for adapted forest management in order to reduce stress factors for forest trees and facilitate a higher ability for regeneration and adaption to climate change.

## Acknowledgments

We thank the responsible institutions of the federal states for providing the data of the forest condition survey of their denser grids. We particularly acknowledge the discussions with Inge Dammann (Northwest German Forest Research Institute), Friedrich Engels (Research Institute for Forest Ecology and Forestry Rhineland-Palatinate), Stefan Meining (Büro für Umweltüberwachung) and Hans Werner Schröck (Research Institute for Forest Ecology and Forestry Rhineland-Palatinate).

# Appendix

```r
library(mgcv)
library(Hmisc)    #for wtd.quantile()
library(MASS)     #for mvrnorm()

# 1) fit spatio-temporal model

#def is defoliation [%]/100; 0 and 1 need to be set a little bit higher and lower, respectively
#x,y are the coordinates
#n_tree is the number of spruce trees per sampling location

data$xy <- factor(paste(data$x, data$y,sep="")) #set up factor for sampling location

mod <- gamm(def~te(y,x,year,bs=c("tp","cr"),d=c(2,1),k=c(25,20))+s(age,bs="cr",k=10),data=data,
correlation=corARMA(form=~year|xy,p=1,q=1),family=gaussian(link="logit"),
weights=data$n_tree, method="REML")

# 2) plot of age effect
int<-mod$gam$coefficient[1]
max_age<-max(data$age)
plot(mod$gam,residuals=FALSE,shade=TRUE,shift=int,trans=function(x)exp(x)/(1+exp(x))*100,
xlim=c(0,max_age),ylim=c(-2,2),las=1,ylab="",xlab="")

# 3) map

mod$gam$data<-data
med_age<-wtd.quantile(data$age[data$year==2015],weights=data$n_tree[data$year==2015],probs=0.5)
```

```r
par(mfrow=c(5,3),oma=c(3,2,0.5,2),mar=c(0.5,1,2,1))
for(i in 1989:2003){
vis.gam(mod$gam,view=c("x","y"),zlim=range(0,0.75),cond=list(year=i,age=med_age),
        n.grid = 60,plot.type="contour",type="response",too.far=0.02,nCol=12,color="terrain",
        main=i,xaxt="n",yaxt="n",ylim=c(5200000,6100000),xlim=c(4000000,4800000))}

# 4) plot of timetrend (age and grid adjusted)

n.sim<- 1000
gobject<- mod$gam
backt <- gobject$family$linkinv
dat <- gobject$data
indi <- order(dat$year)
dat <- dat[indi, ]
pred_age<-med_age

uniq.loc <- unique(dat$xy)
years <- unique(dat$year)
nx <- rep(as.numeric(substring(uniq.loc, 1, 7)), length(years))         # is x coordinate
ny <- rep(as.numeric(substring(uniq.loc, 9, 15)), length(years))        # is y coordinate
nyear <- rep(years, rep(length(uniq.loc), length(years)))
nage <- rep(pred_age, length(nyear))

ndat <- list(nx, ny, nyear, nage)
names(ndat) <- c("x", "y", "year", "age")
ndat <- data.frame(ndat)

M <- predict(gobject, newdata = ndat, type = "lpmatrix")           # re-evaluate smoother basis for new data
                                                                    # and use this as design matrix
```

```
simcoef <- mvrnorm(n = n.sim, coef(gobject), gobject$Vp)          # simulate from predictive distribution

simfit <- as.matrix(M) %*% t(simcoef)

simfit2 <- backt(simfit) * 100                                    #backtransform to response scale

simfit3 <- aggregate(simfit2, by = list(ndat$year), mean)

simfit <- simfit3[, -1]

simquant <- apply(simfit, 1, quantile, p = c(0.025, 0.5, 0.975))  #2.5, 50% und 97.5% quantiles

plot(years, simquant[2, ], type = "p", pch = 19,cex=0.5,col = 1, ylim = c(0, 40), xlab = "",

ylab = "Defoliation[%]", axes = F)

axis(2,las=1); axis(1, at = years, labels = as.character(sort(unique(data$year))))

lines(years, simquant[2, ], lty = 4, col = 1)

lines(years, simquant[1, ], lty = 1, col = 1)

lines(years, simquant[3, ], lty = 1, col = 1)
```

# Table

Table 1 Results of the spatio-temporal modelling and of the simulation study for the grid examination (approach I) presented for the four main tree species. The *p*-value of the spatio-temporal model is valid for both the space-time component (coordinates, year) and the stand age. The federal states Rhineland-Palatinate, Saarland and Schleswig-Holstein were not considered for approach I of the simulation study. $RMPE_{it} > 5\%$ indicates the number of sample plots (counted once only) that had a root mean prediction error of $> 5\%$ defoliation between 2006 and 2015. The number in brackets shows the underlying total number of grid points.

| Tree species | Space-time model $R_{adj}^2$ | *p*-value | Survey grid | $RMPE_t$ [%] Median | Range | $RMPE_{it}$ [%] Median | Range | $RMPE_{it} > 5\%$ |
|---|---|---|---|---|---|---|---|---|
| Spruce | 0.54 | < 0.0001 | 8 x 8 km | 0.6 | 0.5-0.8 | 1.4 | 0.5-6.4 | 0 (745) |
|  |  |  | 16 x 16 km | 1.0 | 0.7-1.3 | 1.8 | 0.6-8.7 | 22 (745) |
| Pine | 0.41 | < 0.0001 | 8 x 8 km | 0.5 | 0.4-1.0 | 1.7 | 0.5-6.1 | 0 (662) |
|  |  |  | 16 x 16 km | 0.9 | 0.5-1.2 | 2.2 | 0.7-9.4 | 76 (662) |
| Beech | 0.47 | < 0.0001 | 8 x 8 km | 0.9 | 0.7-1.1 | 1.7 | 0.6-10.3 | 2 (604) |
|  |  |  | 16 x 16 km | 1.6 | 1.2-1.9 | 2.5 | 0.9-11.7 | 60 (604) |
| Oak | 0.47 | < 0.0001 | 8 x 8 km | 1.1 | 0.8-1.2 | 2.5 | 1.0-9.7 | 14 (381) |
|  |  |  | 16 x 16 km | 1.8 | 1.1-2.4 | 3.3 | 1.3-14.7 | 142 (381) |

**Table 2** Results of the simulation study for the grid examination (approach II) presented for the four main tree species. All federal states were considered for approach II of the simulation study. $RMPE_{it} > 5\%$ indicates the number of sample plots (counted once only) that had a root mean prediction error of $> 5\%$ defoliation between 2006 and 2015. The number in brackets shows the underlying total number of grid points. Results of the spatio-temporal modelling are identical to approach I (see Table 1).

| Tree species | Survey grid | $RMPE_t$ [%] Median | Range | $RMPE_{it}$ [%] Median | Range | $RMPE_{it} > 5\%$ |
|---|---|---|---|---|---|---|
| Spruce | Densification | 0.5 | 0.5-0.8 | 1.4 | 0.5-6.9 | 0 (869) |
|  | 16 x 16 km | 1.0 | 0.8-1.3 | 1.9 | 0.6-8.8 | 20 (869) |
| Pine | Densification | 0.6 | 0.4-1.0 | 1.7 | 0.6-7.0 | 0 (733) |
|  | 16 x 16 km | 0.9 | 0.7-1.2 | 2.3 | 0.7-13.3 | 72 (733) |
| Beech | Densification | 0.8 | 0.6-1.0 | 1.7 | 0.7-12.7 | 0 (775) |
|  | 16 x 16 km | 1.3 | 0.9-1.7 | 2.3 | 0.8-11.9 | 90 (775) |
| Oak | Densification | 1.1 | 0.7-1.5 | 2.4 | 0.7-9.5 | 12 (519) |
|  | 16 x 16 km | 1.9 | 1.3-2.5 | 3.3 | 1.3-14.9 | 166 (519) |

# Figure legends

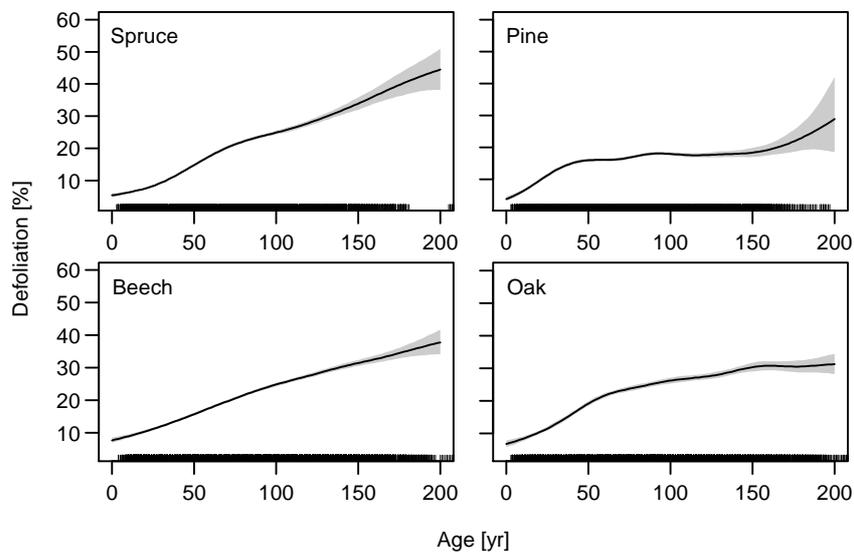

**Figure 1** Effect of the stand age on defoliation of spruce, pine, beech and oak. The lines at the x-axis reflect the observed age values. The grey shaded area indicates the 95% credible interval. Please note that the same scaling was used for the x-axis for reasons of comparability.

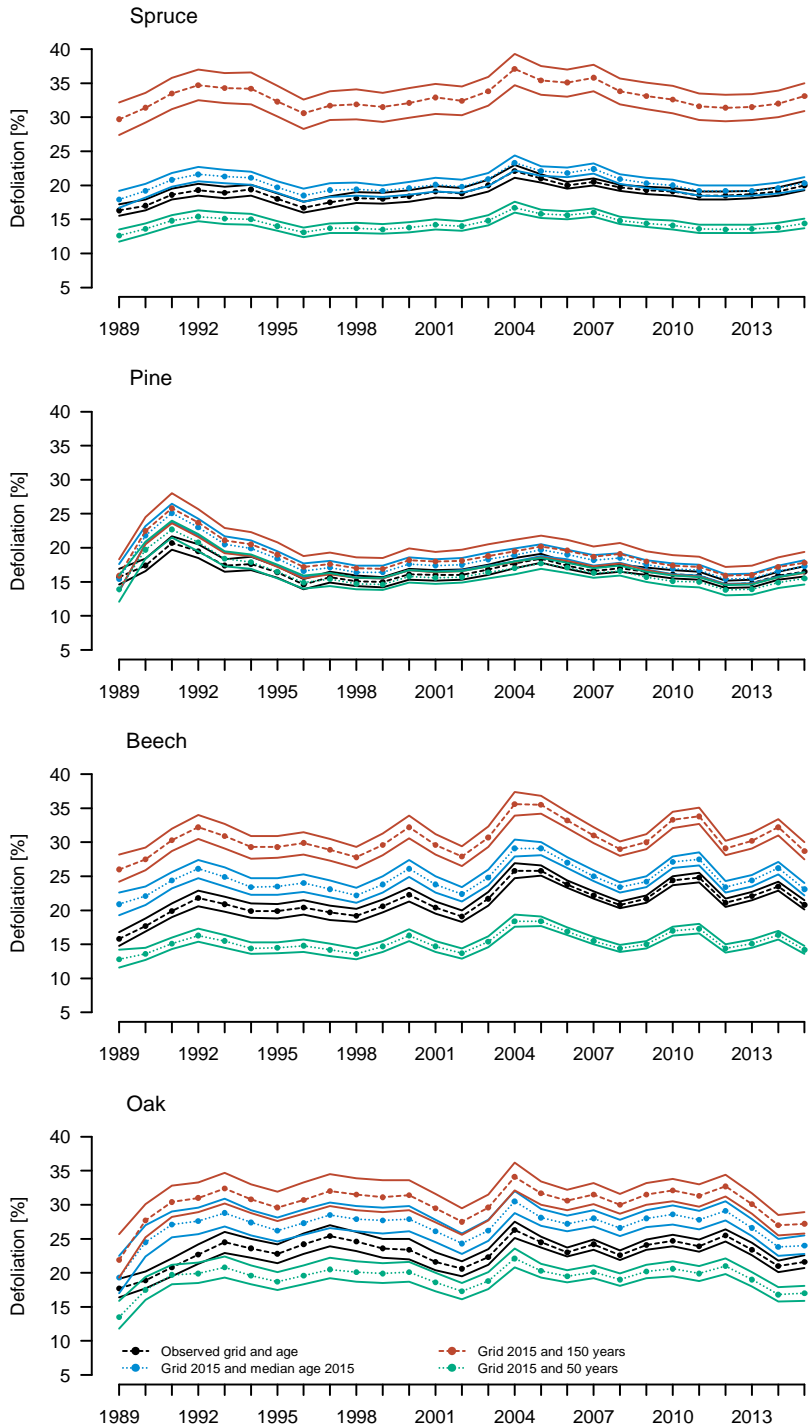

**Figure 2** Estimated mean defoliation and credible interval (2.5% and 97.5% quantiles) for the four main tree species in Germany from 1989 to 2015. In black: Observed 16 x 16 km grid and observed stand age per year (grid 2), in blue: 16 x 16 km grid of 2015 and weighted median stand age of 2015 (spruce: 73 years, pine: 86 years, beech: 104 years, oak: 110 years) (grid 1), in red: 16 x 16 km grid of 2015 and assumed stand age of 150 years, in green: 16 x 16 km grid of 2015 and assumed stand age of 50 years.

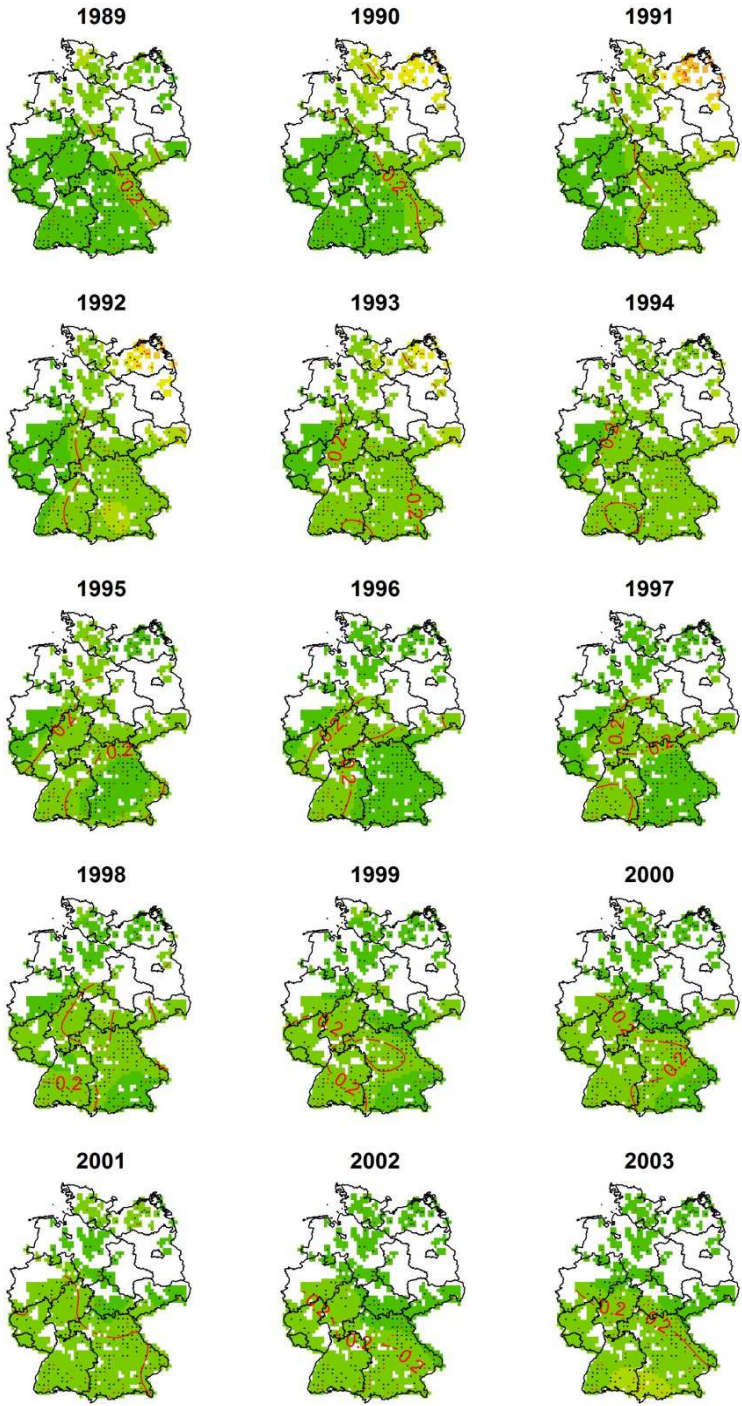

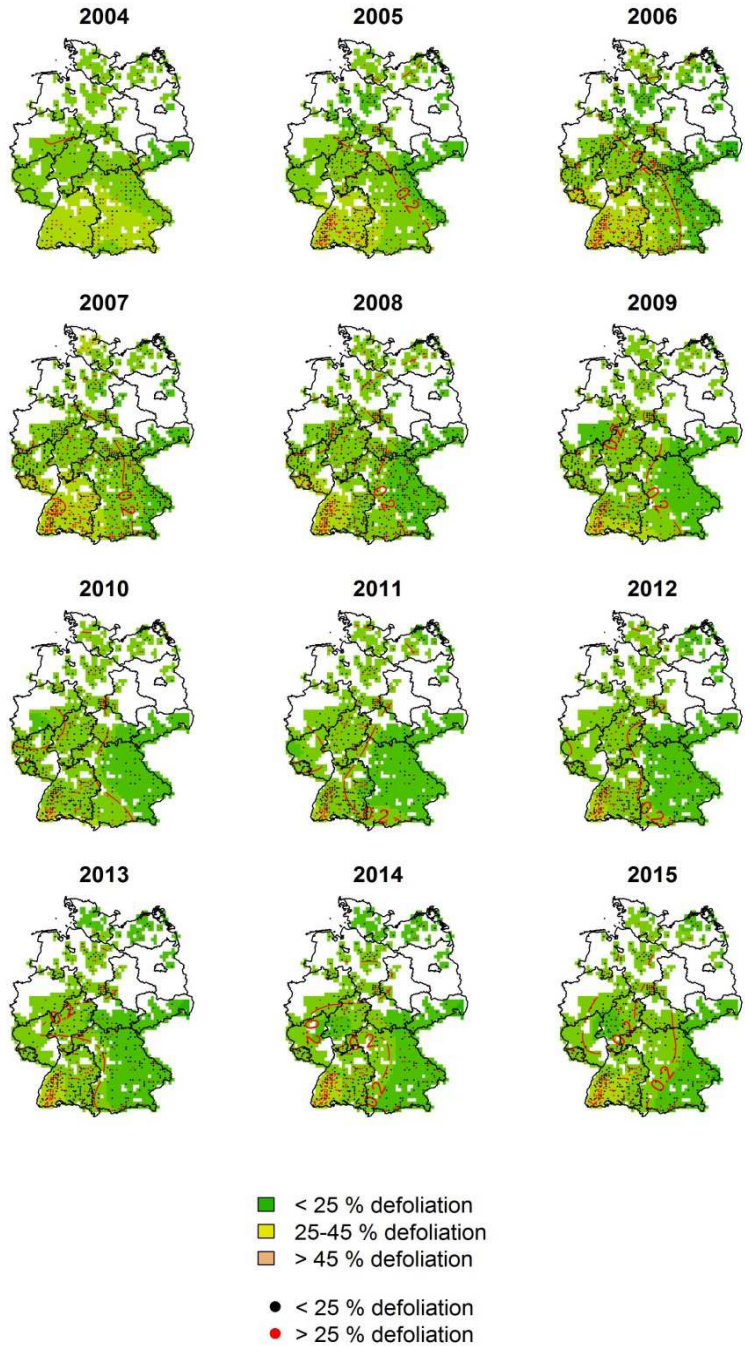

**Figure 3a,b** Results of spatio-temporal modelling of defoliation for spruce from 1989 to 2015 using a nationwide consistent stand age of 73 years. Modelled defoliation is indicated in colour (see legend) and the isolines further reflect the modelled defoliation (e.g. 0.2 is 20%). The sample plots of the respective year are shown as points. Black points indicate plot defoliation < 25% and red points indicate plot defoliation ≥ 25% (observed defoliation at given actual stand age).

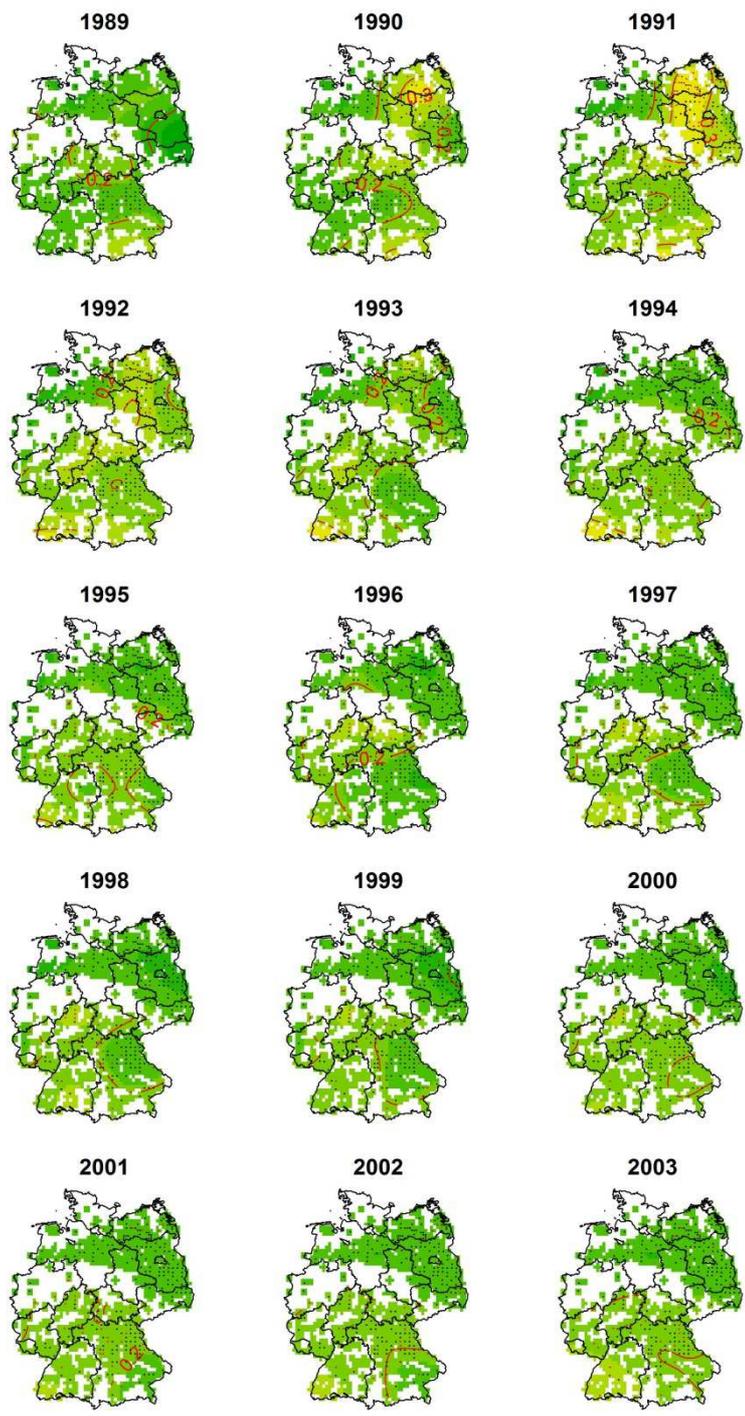

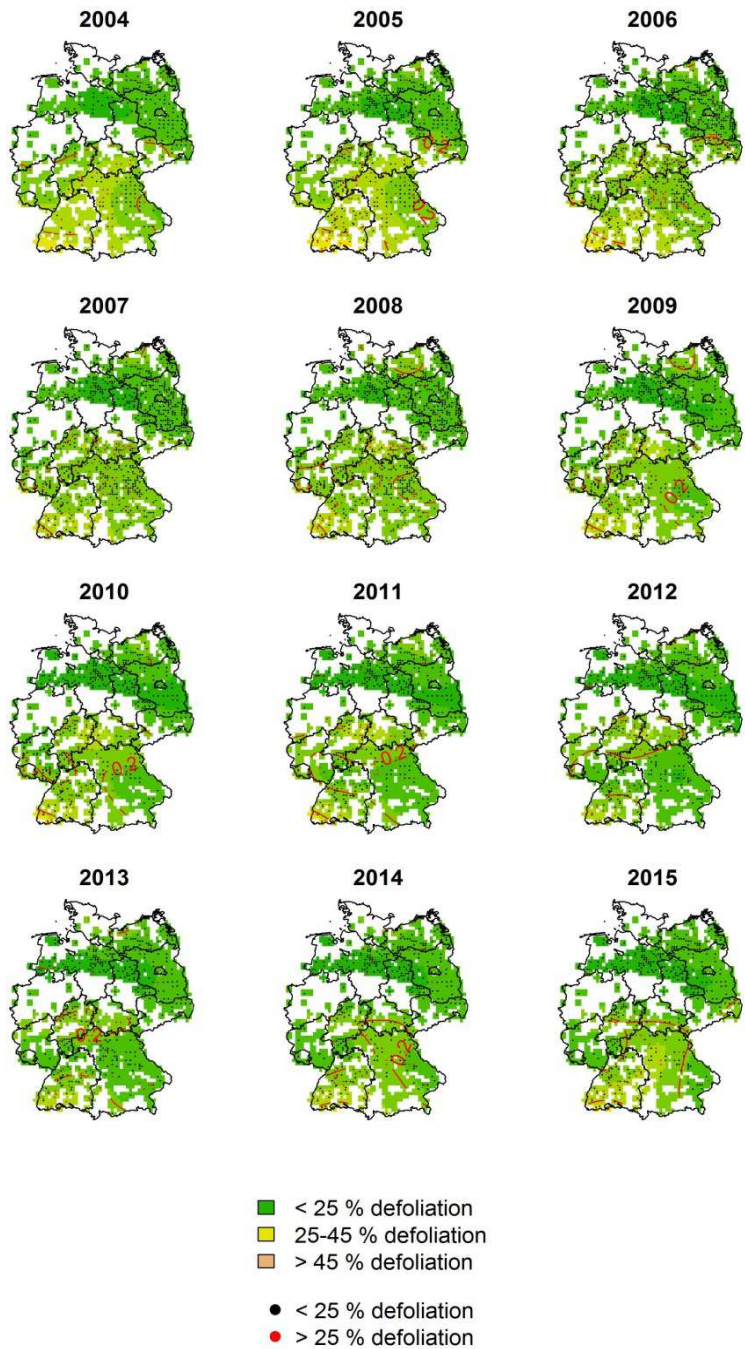

**Figure 4a,b** Results of spatio-temporal modelling of defoliation for pine from 1989 to 2015 using a nationwide consistent stand age of 86 years. Modelled defoliation is indicated in colour (see legend) and the isolines further reflect the modelled defoliation (e.g. 0.2 is 20%). The sample plots of the respective year are shown as points. Black points indicate plot defoliation < 25% and red points indicate plot defoliation ≥ 25% (observed defoliation at given actual stand age).

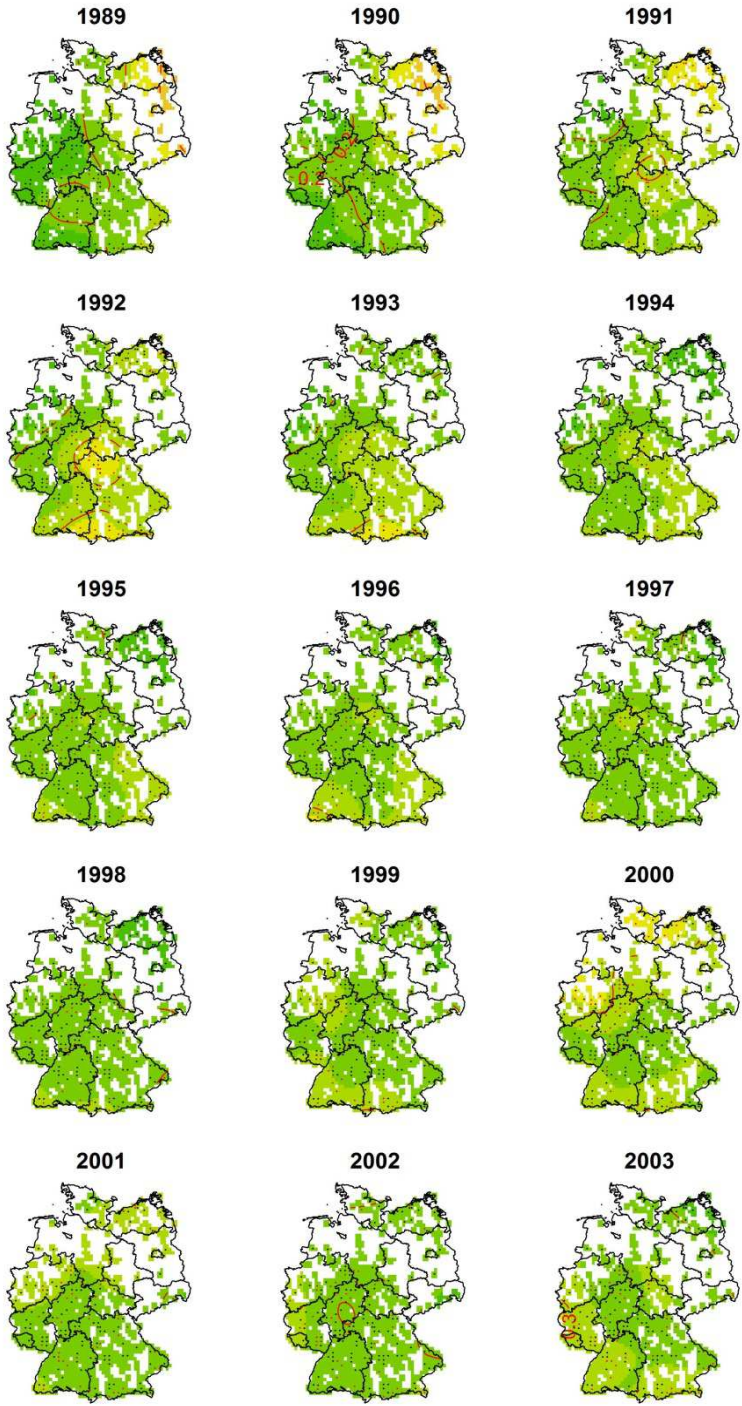

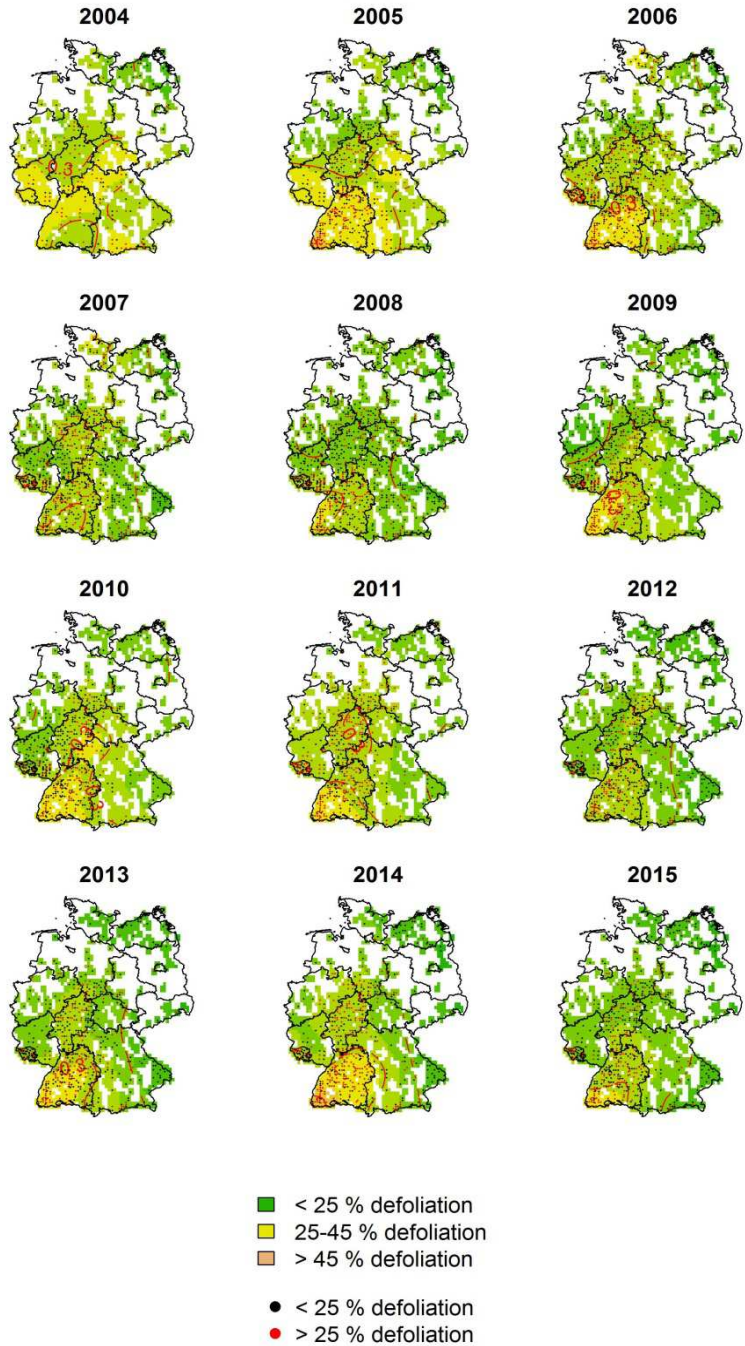

**Figure 5a,b** Results of spatio-temporal modelling of defoliation for beech from 1989 to 2015 using a nationwide consistent stand age of 104 years. Modelled defoliation is indicated in colour (see legend) and the isolines further reflect the modelled defoliation (e.g. 0.2 is 20%). The sample plots of the respective year are shown as points. Black points indicate plot defoliation < 25% and red points indicate plot defoliation ≥ 25% (observed defoliation at given actual stand age).

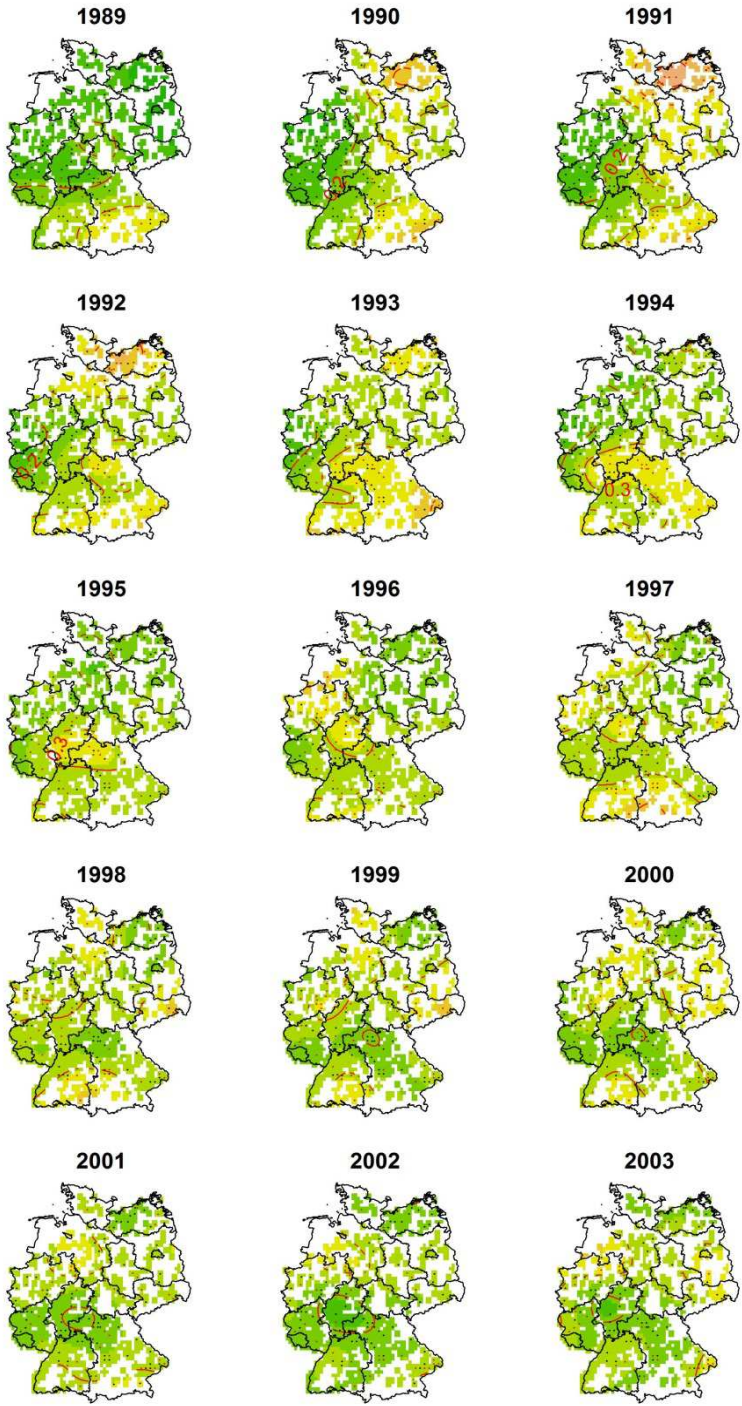

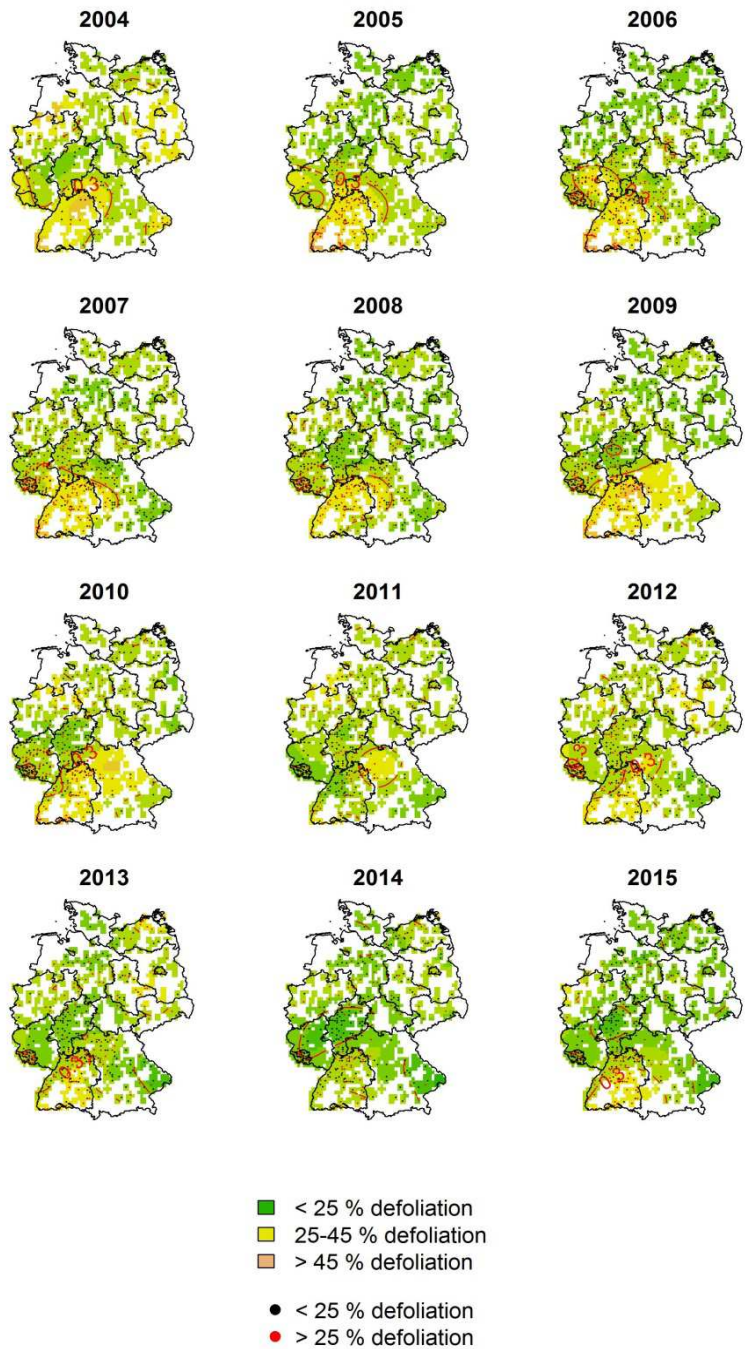

**Figure 6a,b** Results of spatio-temporal modelling of defoliation for oak from 1989 to 2015 using a nationwide consistent stand age of 110 years. Modelled defoliation is indicated in colour (see legend) and the isolines further reflect the modelled defoliation (e.g. 0.2 is 20%). The sample plots of the respective year are shown as points. Black points indicate plot defoliation < 25% and red points indicate plot defoliation ≥ 25% (observed defoliation at given actual stand age).

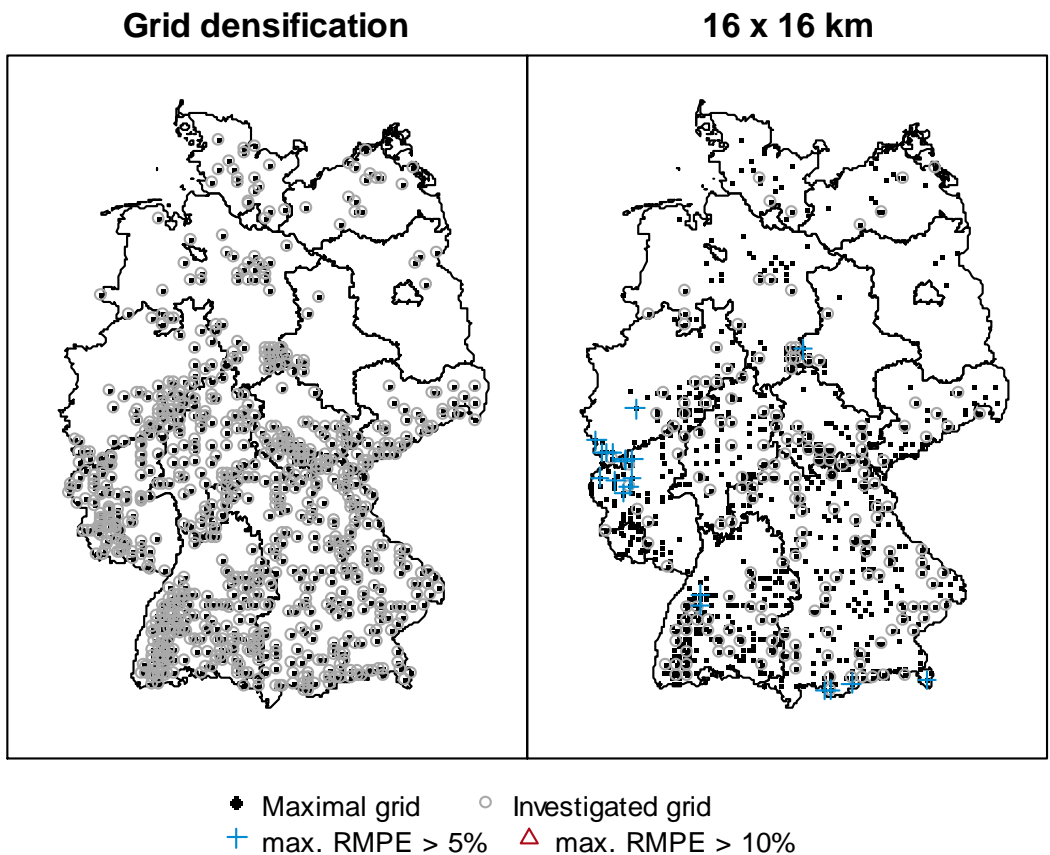

**Figure 7** Highest estimated root mean prediction error (RMPE$_{it}$) of defoliation of spruce for the grid densification (at least 8 x 8 km grid) and 16 x 16 km grid (approach II) for the years 2006 to 2015. Black points indicate all possible sample plots (corresponds to the grid densification). The grey circles show the sample plots considered for the estimation of the RMPE (grid densification and 16 x 16 km grid, respectively). Blue crosses indicate RMPE$_{it}$ > 5% and red triangles RMPE$_{it}$ > 10%. The federal states Rhineland-Palatinate, Saarland and Schleswig-Holstein have deviating grid densities from 8 x 8 km (denser grids).

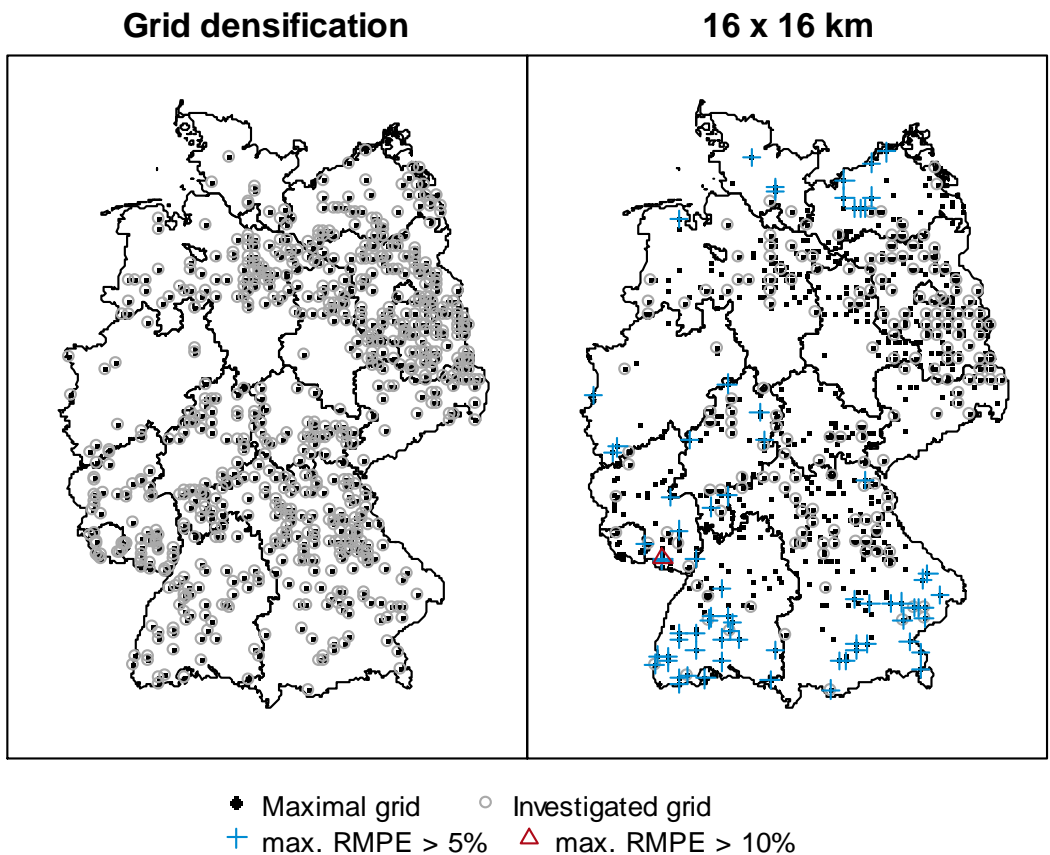

**Figure 8** Highest estimated root mean prediction error (RMPE$_{it}$) of defoliation of pine for the grid densification (at least 8 x 8 km grid) and 16 x 16 km grid (approach II) for the years 2006 to 2015. Black points indicate all possible sample plots (corresponds to the grid densification). The grey circles show the sample plots considered for the estimation of the RMPE (grid densification and 16 x 16 km grid, respectively). Blue crosses indicate RMPE$_{it}$ > 5% and red triangles RMPE$_{it}$ > 10%. The federal states Rhineland-Palatinate, Saarland and Schleswig-Holstein have deviating grid densities from 8 x 8 km (denser grids).

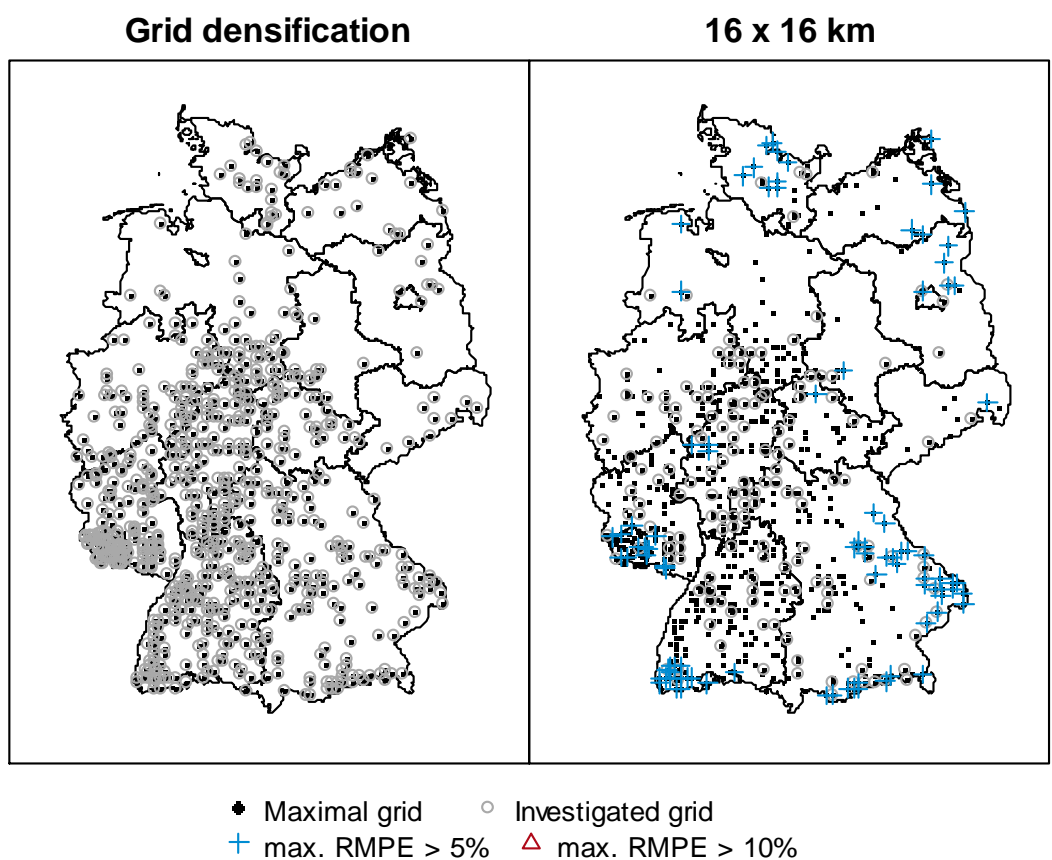

**Figure 9** Highest estimated root mean prediction error (RMPE$_{it}$) of defoliation of beech for the grid densification (at least 8 x 8 km grid) and 16 x 16 km grid (approach II) for the years 2006 to 2015. Black points indicate all possible sample plots (corresponds to the grid densification). The grey circles show the sample plots considered for the estimation of the RMPE (grid densification and 16 x 16 km grid, respectively). Blue crosses indicate RMPE$_{it}$ > 5% and red triangles RMPE$_{it}$ > 10%. The federal states Rhineland-Palatinate, Saarland and Schleswig-Holstein have deviating grid densities from 8 x 8 km (denser grids).

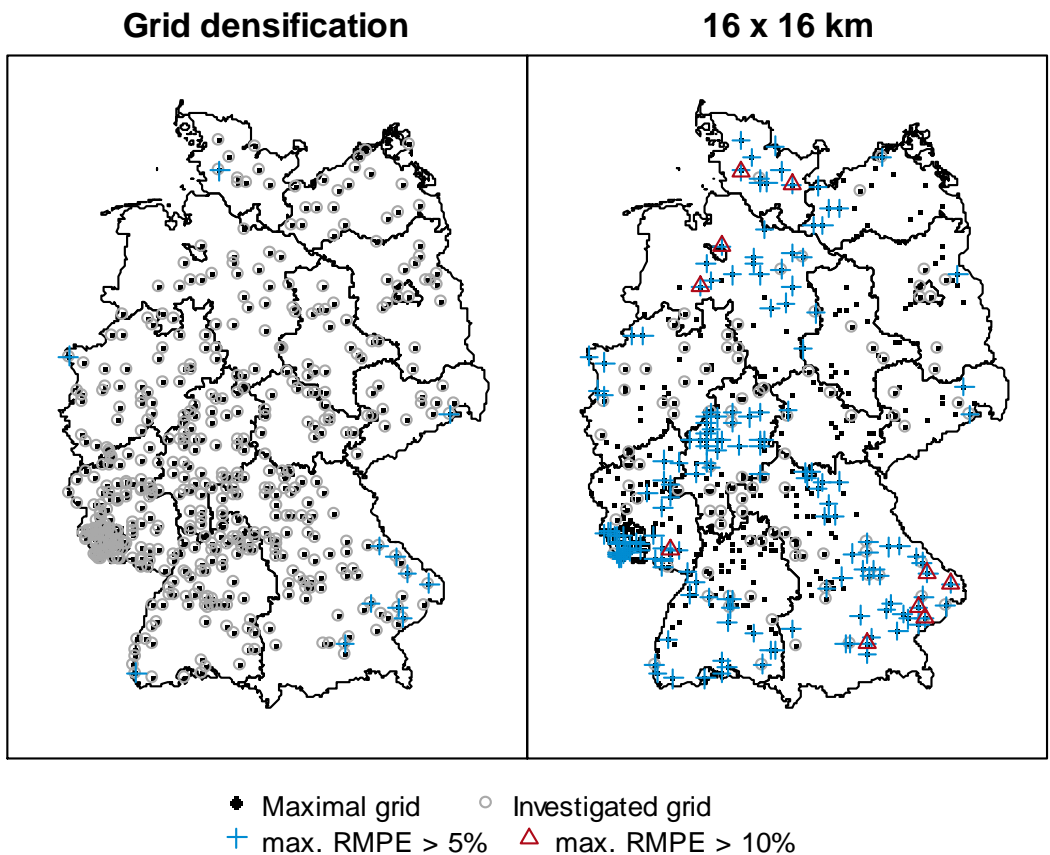

**Figure 10** Highest estimated root mean prediction error ($RMPE_{it}$) of defoliation of oak for the grid densification (at least 8 x 8 km grid) and 16 x 16 km grid (approach II) for the years 2006 to 2015. Black points indicate all possible sample plots (corresponds to the grid densification). The grey circles show the sample plots considered for the estimation of the RMPE (grid densification and 16 x 16 km grid, respectively). Blue crosses indicate $RMPE_{it} > 5\%$ and red triangles $RMPE_{it} > 10\%$. The federal states Rhineland-Palatinate, Saarland and Schleswig-Holstein have deviating grid densities from 8 x 8 km (denser grids).